\documentclass{article}

\usepackage{amsmath,amssymb,amsfonts}
\usepackage{booktabs}
\usepackage{graphicx}
\usepackage{multirow}
\usepackage{hyperref}
\usepackage[a4paper, total={6in, 10in}]{geometry}
\usepackage[backend=biber,style=numeric]{biblatex}

\AtBeginDocument{
  }

\begin{document}
\title{Detection of LLM-assisted Code Plagiarism Using k-gram Software Birthmarks}
\date{}

\author{
Nikolay Fedorov
\thanks{Corresponding author. N. Fedorov is with Graduate School of Environmental, Life, Natural Science and Technology, Okayama University, Okayama, Japan}
\\ \href{mailto:fedorov_n@s.okayama-u.ac.jp}{fedorov\_n@s.okayama-u.ac.jp}
\and
Akito Monden
\thanks{A. Monden is with Faculty of Environmental, Life, Natural Science and Technology, Okayama University, Okayama, Japan}
\\ \href{mailto:monden@okayama-u.ac.jp}{monden@okayama-u.ac.jp}
\and
Hiroki Inayoshi
\thanks{H. Inayoshi is with Faculty of Environmental, Life, Natural Science and Technology, Okayama University, Okayama, Japan}
\\ \href{mailto:inayoshi@okayama-u.ac.jp}{inayoshi@okayama-u.ac.jp} 
\and
Haruaki Tamada
\thanks{H. Tamada is with Faculty of Information Science and Engineering, Kyoto Sangyo University, Kyoto, Japan}
\\ \href{mailto:tamada@cc.kyoto-su.ac.jp}{tamada@cc.kyoto-su.ac.jp} 
\and
Masateru Tsunoda
\thanks{M. Tsunoda is with Faculty of Informatics, Cyber Informatics Research Institute, Kindai University, Higashiosaka-shi, Japan}
\\ \href{mailto:tsunoda@info.kindai.ac.jp}{tsunoda@info.kindai.ac.jp}
}

\maketitle

\begin{abstract}

Large language models (LLMs) have significantly lowered the technical barrier to software plagiarism. 
By transforming existing source code while preserving its functionality, modern LLMs can generate semantically identical program that may evade traditional plagiarism detection techniques.

Among such attacks, {\it code paraphrasing} modifies the syntax and structure of a program while preserving its behavior. 
This paper investigates whether software birthmarks can detect such LLM-assisted plagiarism. 
As a starting point, we employ k-gram software birthmarks based on unique k-grams of Java opcodes, with k ranging from 1 to 6.

We employ three contemporary LLMs: ChatGPT-5.1-Codex-Mini, DeepSeek-V4-Flash, and Claude-Haiku-4.5. 
The dataset consists of individually compilable source files extracted from actively maintained BSD-2-Clause licensed Java projects. 
We further compare five similarity measures for birthmark matching: cosine similarity, Dice index, Jaccard coefficient, Simpson index, and edit-distance-based similarity.

The results demonstrate that k-gram software birthmarks remain effective for detecting LLM-assisted plagiarism. 
Among the evaluated models, ChatGPT-5.1-Codex-Mini generated the most difficult-to-detect clones.
Furthermore, the findings confirm the higher performance of coding-oriented models for plagiarism task.

\medskip

\end{abstract}

\maketitle

\section{Introduction}
\label{sec:intro}
Recent advances in large language models (LLMs) and AI agents have transformed software development. 
Modern LLMs are capable of generating, modifying, and explaining source code with a level of quality that was previously achievable only by experienced developers
~\cite{dong2025surveycodegenerationllmbased, guo2024deepseekcoderlargelanguagemodel, Brach2024canLLMdetectplagiarism}.
While these capabilities improve developer productivity, they also create new opportunities for software plagiarism~\cite{PARK2025112454, Brach2024canLLMdetectplagiarism, he2025execoderempoweringlargelanguage}.

In particular, an LLM can be instructed to rewrite an existing program while preserving its functionality, a process often referred to as \emph{code paraphrasing}~\cite{PARK2025112454}. 
By modifying identifiers, control structures, formatting styles, and other syntactic characteristics, LLMs can generate code that appears substantially different from the original implementation while maintaining equivalent behavior.

Such capabilities pose a significant threat to open-source software. Open-source projects are typically distributed under licenses that regulate software reuse and redistribution. In particular, copyleft licenses such as the GNU General Public License (GPL)\footnote{\url{https://www.gnu.org/licenses/gpl-3.0.en.html}} require derivative software to be distributed under the same license. 
Traditionally, reproducing the functionality of an existing project without directly copying its code required considerable software engineering expertise~\cite{romansky2018sourcerersapprenticestudycode}. 
However, LLM-based code paraphrasing significantly lowers this barrier, enabling users to generate functionally equivalent implementations with substantially reduced effort~\cite{Brach2024canLLMdetectplagiarism}. 
Such practices may undermine the collaborative ecosystem on which open-source software development depends.

Many existing plagiarism detection techniques focus on source-code similarity and code-clone detection~\cite{romansky2018sourcerersapprenticestudycode, Le2013CodeSimilarityDetectionTool,Golubev2020, Brach2024canLLMdetectplagiarism}. 
While such approaches are effective in educational settings where source code submissions are available, real-world software plagiarism often involves the redistribution of executable artifacts rather than source code. 
Consequently, techniques that can operate on compiled software are required.

Among the various plagiarism detection approaches, {\it software birthmarks} provide an attractive solution.
A software birthmark is a set of characteristic features extracted from a program that enables discrimination between plagiarized and independently developed software~\cite{Tamada2004DesignAE, Tamada2005, Nazir2019}. 
Because birthmarks can be extracted from executable files, they are particularly attractive for practical plagiarism detection scenarios~\cite{Tamada2007DynamicAPIBirthmarks, Tamada2004DynamicBirthmarksWindows, Lu2007}. 
In this study, we employ the k-gram birthmark, which represents software using unique k-grams of Java bytecode instructions~\cite{Myles2005, Lee2016APIBasedSB}.

The objective of this work is to investigate whether k-gram software birthmarks remain effective against LLM-assisted software plagiarism via code paraphrasing. 
Using three contemporary LLMs -- ChatGPT-5.1-Codex-Mini\footnote{\url{https://developers.openai.com/api/docs/models/gpt-5.1-codex-mini}}, DeepSeek-V4-Flash\footnote{\url{https://api-docs.deepseek.com/news/news260424}}, and Claude-Haiku-4.5\footnote{\url{https://www.anthropic.com/news/claude-haiku-4-5}} -- we generate plagiarized versions of Java source files and evaluate the effectiveness of k-gram birthmarks with multiple similarity measures.

The contributions of this paper are summarized as follows:

\begin{itemize}
\item We investigate the effectiveness of k-gram software birthmarks for detecting LLM-assisted software plagiarism.

\item We conduct an empirical evaluation using three state-of-the-art LLMs and compare multiple similarity functions and k-gram configurations.

\item We analyze the relative difficulty of detecting plagiarism generated by different LLMs for code paraphrasing scenario.

\end{itemize}

The remainder of this paper is organized as follows. 
Section~\ref{sec:birthmark} introduces software birthmarks and k-gram birthmarks. 
Section~\ref{sec:llm-plagiarism} describes LLM-assisted plagiarism and code paraphrasing. 
Section~\ref{sec:experiment} presents the experimental design and dataset preparation. 
Section~\ref{sec:results} reports and discusses the experimental results. 
Finally, Section~\ref{sec:conclusion} concludes the paper and outlines future work.

\section{Software Birthmarks}
\label{sec:birthmark}

The concept of a software birthmark was introduced by Tamada et al.~\cite{Tamada2004DesignAE,Tamada2007DynamicAPIBirthmarks}. 
A software birthmark is a set of characteristic features extracted from a program that can be used to distinguish plagiarized software from independently developed software. 
Unlike source-code-based plagiarism detection techniques, software birthmarks can be extracted from executable artifacts, making them suitable for practical plagiarism detection scenarios in which source code is unavailable.

A useful software birthmark should satisfy two important properties: \emph{resilience} and \emph{credibility}~\cite{Lee2016APIBasedSB,Bai2008}. 
Resilience refers to the ability of a birthmark to remain recognizable after plagiarism-related transformations. 
Let $P$ be an original program and $Q$ be a plagiarized version of $P$. 
A resilient birthmark produces similar representations for both programs despite modifications introduced during the plagiarism process.

Credibility refers to the ability of a birthmark to distinguish independently developed software. 
Let $P$ and $Q$ be two independently developed programs. 
A credible birthmark should produce sufficiently different representations for these programs, thereby avoiding false plagiarism detections.

In practice, a birthmark extraction technique defines a function $B(\cdot)$ that maps a program to its birthmark representation. 
The similarity between two programs is then estimated by comparing their birthmarks using an appropriate similarity function. 
The effectiveness of a software birthmark therefore depends both on the extracted representation and on the similarity measure used to compare birthmarks.

\subsection{k-gram Birthmark}
\label{sec:k-gram-birthmark}

In this study, we employ the k-gram birthmark proposed by Myles and Collberg~\cite{Myles2005}. 
The k-gram birthmark represents a program as a set of unique contiguous sequences of $k$ bytecode instructions (opcodes). 
Because it is extracted from compiled code, the birthmark can be used even when source code is unavailable.

A k-gram is constructed by sliding a window of length $k$ over a sequence of bytecode instructions. For example, given the instruction sequence

\begin{center}
\texttt{aload\_0, getfield, ifnull, return},
\end{center}

the corresponding 2-grams are

\begin{center}
\texttt{(aload\_0, getfield)},\
\texttt{(getfield, ifnull)},\
\texttt{(ifnull, return)},
\end{center}

or, if converting to opcode numbers

\begin{center}
\texttt{(25, 180)},\
\texttt{(180, 198)},\
\texttt{(198, 177)}.
\end{center}

The birthmark of a program is defined as the set of unique k-grams extracted from its bytecode representation. 
Similar programs are expected to share a larger number of k-grams than independently developed programs.

The k-gram birthmark was selected for this study for three reasons. 
First, it can be extracted directly from executable artifacts, making it suitable for practical plagiarism detection scenarios. 
Second, it has been extensively evaluated in previous software birthmark studies~\cite{Myles2005,Lee2016APIBasedSB}. 
Third, its simplicity allows us to focus on the impact of LLM-based code transformations rather than on the complexity of the birthmark extraction process itself.

For each source file in our dataset, we first compile the program using the Java compiler and extract bytecode instructions using the \texttt{javap} tool. 
The extracted instructions are then converted into opcode representations using the opcode definitions provided by the ASM framework\footnote{\url{https://gitlab.ow2.org/asm/asm}}. 
Finally, unique k-grams are generated from the resulting opcode sequences.

Following previous studies on k-gram birthmarks, we evaluate values of $k$ ranging from 1 to 6. 
Smaller values of $k$ generally provide greater robustness against code transformations, whereas larger values capture more structural information but may become more sensitive to modifications introduced during plagiarism\cite{Myles2005}. 
By evaluating multiple values of $k$, we investigate the trade-off between robustness and descriptiveness in the context of LLM-based code paraphrasing.

\subsection{Similarity Functions}
\label{sec:sim-functions}

The effectiveness of a software birthmark depends not only on the extracted representation but also on the method used to compare birthmarks. 
Different similarity functions emphasize different characteristics of the compared data and may therefore exhibit different levels of effectiveness when detecting plagiarized software. 
Consequently, this study evaluates multiple similarity functions in order to investigate their suitability for detecting LLM-generated code clones.

Following our previous studies~\cite{Fedorov2024,fedorov2026projectwisecomparisonsoftwarebirthmark}, we employ the following similarity functions:

\begin{itemize}
\item Cosine similarity using count vectorization;
\item Cosine similarity using TF-IDF vectorization;
\item Dice--S{\o}rensen coefficient (Dice index);
\item Jaccard similarity coefficient;
\item Simpson similarity index;
\item Edit-distance-based similarity using the Levenshtein distance.
\end{itemize}

These functions represent three different approaches to similarity calculation.

Vector-based methods represent birthmarks as feature vectors and compare them using the angle between vectors. In this study, we evaluate cosine similarity with both count-based and TF-IDF-based vectorization schemes.

Set-based methods compare birthmarks based on the overlap between sets of extracted k-grams. We employ three widely used set similarity measures: the Dice index, Jaccard coefficient, and Simpson index.

Finally, sequence-based methods take the ordering of elements into account. To represent this category, we employ an edit-distance-based similarity measure derived from the Levenshtein distance.

By evaluating representative functions from these three categories, we aim to investigate how the choice of similarity measure affects the effectiveness of k-gram birthmarks for detecting LLM-assisted code plagiarism.

\section{LLM-assisted Code Plagiarism}
\label{sec:llm-plagiarism}

The emergence of large language models (LLMs) has fundamentally changed the landscape of software plagiarism. 
Traditionally, reproducing the functionality of an existing software system while concealing direct code reuse required substantial programming expertise and development effort~\cite{romansky2018sourcerersapprenticestudycode}. 
Modern LLMs can significantly reduce this effort by automatically transforming source code while preserving its behavior~\cite{PARK2025112454}.

As a result, software plagiarism is no longer limited to direct copying of source code. 
An attacker may use an LLM to modify the lexical, syntactic, and structural characteristics of a program while maintaining its functionality. 
Such transformations can potentially reduce the effectiveness of traditional plagiarism detection techniques that rely on source-code similarity~\cite{Le2013CodeSimilarityDetectionTool,Golubev2020, Brach2024canLLMdetectplagiarism}.

A representative example of this threat is \emph{code paraphrasing}, which was recently investigated by Park et al.~\cite{PARK2025112454}. 
In this approach, a source code fragment together with a paraphrasing instruction is provided to an LLM, which generates a modified version of the program while attempting to preserve its original functionality. 
The objective of code paraphrasing is not to introduce new functionality but rather to alter the appearance of the program. 
Typical modifications include changes to identifier names, code formatting, control structures, statement organization, and other syntactic elements.
Consequently, the generated program may differ substantially from the original source code while exhibiting equivalent behavior.

Park et al.~\cite{PARK2025112454} demonstrated that modern LLMs are capable of generating functionally equivalent code with reduced source-code similarity, suggesting that LLM-assisted transformations may complicate traditional plagiarism detection. 
Motivated by these findings, this work investigates whether software birthmarks extracted from executable artifacts remain effective against such transformations.

To generate plagiarized software samples, we adopt the code paraphrasing approach of Park et al. with several modifications. First, we provide more explicit instructions regarding the required transformations. 
Instead of requesting general modifications to reduce similarity, we explicitly instruct the model to modify the syntax, formatting, and structure of the program while preserving its functionality. 
Second, we require the generated code to remain compilable using the Java compiler and instruct the model to avoid introducing additional dependencies. 
These modifications were introduced to increase the consistency of the generated outputs and to ensure that the resulting programs can be processed by the subsequent birthmark extraction pipeline.

The complete prompt used in this study is provided in Appendix~\ref{sec:used-prompts}.

\section{Experiment Design}
\label{sec:experiment}

\subsection{Research Questions}

This study investigates the effectiveness of k-gram software birthmarks for detecting plagiarism generated through LLM-based code paraphrasing. To guide the evaluation, we define the following research questions:

\begin{itemize}
\item \textbf{RQ1:} How does the choice of k affect the effectiveness of plagiarism detection?

\item \textbf{RQ2:} Which similarity function provides the best performance for detecting LLM-generated code clones?

\item \textbf{RQ3:} How does the choice of LLM affect the detectability of generated code clones?

\end{itemize}

\subsection{Selected LLMs}

To evaluate the impact of different LLMs on plagiarism detectability, we employ three contemporary models from different providers:

\begin{itemize}
\item ChatGPT-5.1-Codex-Mini by OpenAI;
\item DeepSeek-V4-Flash by DeepSeek;
\item Claude-Haiku-4.5 by Anthropic.
\end{itemize}

These models were selected because they are publicly accessible and represent different approaches to code generation and transformation. By evaluating multiple providers, we aim to reduce the risk that the results are specific to a single LLM family.

\subsection{Dataset Preparation}
\label{sec:dataset}

The dataset was prepared using the following approach:
\begin{itemize}
    \item Initial repository collection: 
    We use open-source repositories published on GitHub\footnote{\url{https://github.com/}} under the BSD-2-Clause license\footnote{\url{https://opensource.org/license/BSD-2-Clause}}.
    We take relatively popular projects (interpreted as having over 100 stars) and actively maintained (being pushed at least once since June 14th, 2025).
    The repositories were collected on June 14th, 2026. 
    \item Standalone compilable source file filtering: 
    Because in this work we use k-gram birthmark we focus on compiled files instead of the source code. 
    Additionally, we expect that malicious actor is unlikely to publish the plagiarized source code to the public, therefore, focusing on the executable file comparison is a more likely scenario to a real-world case. 
    Finally, because the focus of this work is on the module-wise plagiarism, we select only the source file that can be compiled individually using the Java compiler. 
    \item Logical line count (LLC) filtering: 
    Finally, to exclude small files such as basic classes or interfaces, we use a filtering method proposed in out previous work~\cite{fedorov2026projectwisecomparisonsoftwarebirthmark} which ignores comment or newline lines and instead focuses on actual lines of code of a given source file, denoted as logical line count (LLC).
    In this work, we only select the source files with over 30 such lines.
    In addition, some repositories contained large configuration files that stored constant values used in the project. 
    Because cloning such files does not hold any advantage and due to the limited context window of the models, files with LLC of over 1000 were excluded. 
\end{itemize}

Source file count for each stage of dataset data collection is provided on Table~\ref{tab:source-file-count}. 
\begin{table}
    \centering
    \small
    \caption{The number of source files.}
    \label{tab:source-file-count}
    \begin{tabular}{c|c|c|c}
    \toprule
        & Initial & After compilable-file filtering & After LLC filtering \\
        \midrule
        Source file count & 19277 & 1812 & 352 \\
        \bottomrule
    \end{tabular}
\end{table}

\subsection{Evaluation Procedure}
\label{sec:evaluation-procedure}

For each original source file, a paraphrased version generated by an LLM is compared against the birthmark extracted from the original file (using the prompt shown in Appendix~\ref{sec:used-prompts}.)
These comparisons are used to evaluate the \emph{resilience} of the birthmark, i.e., its ability to identify plagiarized software despite modifications introduced during paraphrasing.

To evaluate \emph{credibility}, each original birthmark is additionally compared against an equal number of birthmarks extracted from unrelated source files belonging to different projects. 
This ensures a balanced evaluation of plagiarism and non-plagiarism cases. The procedure is repeated for every combination of LLM, k-gram configuration, and similarity function.

Following our previous work~\cite{fedorov2026projectwisecomparisonsoftwarebirthmark}, we evaluate detection performance using the harmonic mean (Hmean) of resilience and credibility rates. Resilience rate and credibility rate are defined as

\begin{equation}
\text{resilience rate}
=
\frac{TP}{TP + FN},
\end{equation}

\begin{equation}
\text{credibility rate}
=
\frac{TN}{TN + FP}.
\end{equation}

where TP, TN, FP, and FN denote true positives, true negatives, false positives, and false negatives, respectively.

Resilience rate measures the ability of a birthmark to correctly identify plagiarized software, whereas credibility rate measures its ability to correctly distinguish independently developed software. The harmonic mean of these two quantities is calculated as

\begin{equation}
\text{Hmean}
=
\frac{2}{
\frac{1}{\text{resilience rate}}
+
\frac{1}{\text{credibility rate}}
}.
\end{equation}

Hmean provides a balanced evaluation by assigning equal importance to resilience and credibility. Consequently, a high Hmean score can only be achieved when both plagiarism detection and false-positive avoidance perform well.

Software birthmark similarity is evaluated using the threshold-based decision rule described as follows:
\begin{equation}
    Sim(X,Y)  
    \begin{cases}
            \leq 1 - \varepsilon & X \nsim Y \\
            > \varepsilon & X \sim Y \\
            \text{inconclusive}
    \end{cases}
\label{eq:birthmark-sim-eval}
\end{equation}
where $Sim(X,Y)$ is the similarity between $X$ and $Y$, and $\varepsilon$ ($0.5 \leq \varepsilon < 1$) is a threshold.

Because the available dataset is relatively small, we do not divide the data into separate training and testing subsets. 
Instead, for each experimental configuration, we report the best achievable Hmean obtained from the calculated similarity values. This approach allows us to compare the relative effectiveness of different k-gram configurations, similarity functions, and LLMs under identical conditions.

\subsection{Compile Errors in Paraphrased Code}
\label{sec:compile-errors}

Although the dataset was constructed from individually compilable Java source files and the paraphrasing prompt explicitly instructed the LLMs to generate compilable code, some generated outputs still contained compilation errors. 
Such errors included syntax mistakes, unresolved references, and modifications that violated Java compilation rules.

Because k-gram birthmarks are extracted from Java bytecode, successfully compiling the paraphrased code is a prerequisite for subsequent analysis. Therefore, paraphrased files that could not be compiled were excluded from the experiment.

Table~\ref{tab} summarizes the number of compilation failures observed for each evaluated model. 
ChatGPT-5.1-Codex-Mini produced the smallest number of compilation errors, whereas DeepSeek-V4-Flash and Claude-Haiku-4.5 generated a larger number of non-compilable outputs. 
Nevertheless, the overall compilation success rate remained high for all evaluated models, allowing the remaining paraphrased files to be used in the subsequent birthmark evaluation.

\begin{table}
\centering
\small
\caption{Compile errors in paraphrased code.}
\label{tab}
\begin{tabular}{c|c|c|c}
\toprule 
& ChatGPT-5.1-Codex-Mini & DeepSeek-V4-Flash & Claude-Haiku-4.5  \\
\midrule
Compile errors & 10 & 39 & 54\\
\bottomrule
\end{tabular}
\end{table}

\section{Experimental Results}
\label{sec:results}

\subsection{RQ1: Impact of \texorpdfstring{$k$}{k} on Detection Performance}

\begin{figure*}
    \centering
    \includegraphics[width=0.8\linewidth]{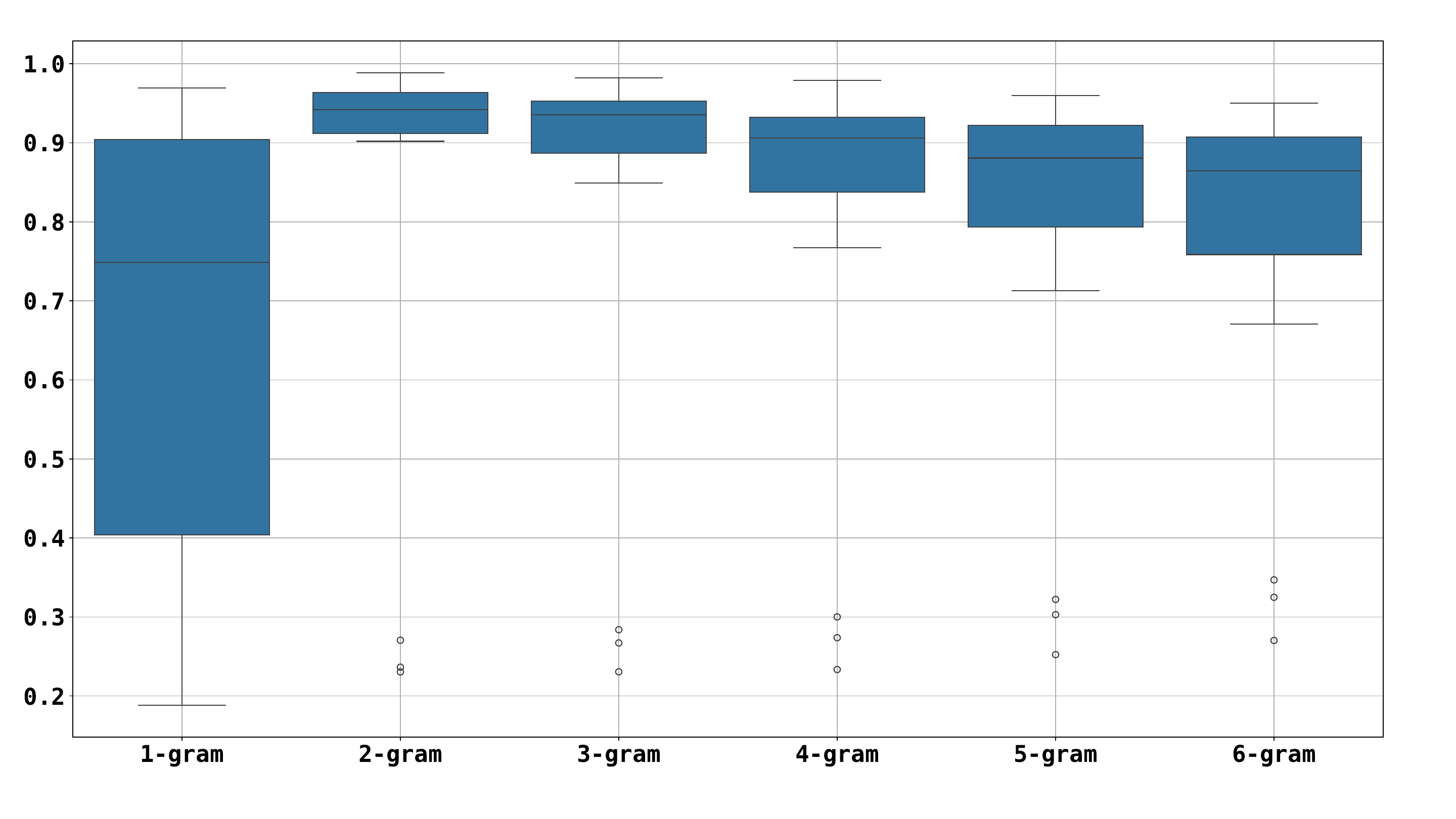}
    \caption{Detection performance of k-gram birthmarks against LLM-generated code clones in terms of Hmean.}
    \label{fig:birthmark-comp}
\end{figure*}

Figure~\ref{fig:birthmark-comp} presents the overall detection performance of the evaluated k-gram birthmarks. 
The results reveal a clear relationship between the value of $k$ and the effectiveness of plagiarism detection.

Among all evaluated birthmarks, 1-gram consistently achieved the lowest performance. 
Because a 1-gram birthmark represents software using only individual opcodes, it captures very limited structural information. 
Consequently, many unrelated programs share similar opcode distributions, reducing the credibility of the birthmark and leading to lower overall Hmean values.
Also, the results show that the 2-gram birthmark achieved the highest average performance (0.830), closely followed by the 3-gram birthmark (0.822). 
For larger values of $k$, detection performance gradually decreased.

This trend suggests that short opcode sequences provide an effective balance between resilience and credibility for detecting LLM-generated code clones. 
While increasing $k$ allows the birthmark to capture more structural information, longer opcode sequences are also more sensitive to modifications introduced during code paraphrasing. 
As LLMs frequently alter control structures, statement organization, and other implementation details, long opcode patterns are less likely to remain unchanged between the original and paraphrased versions.

Overall, the results indicate that small values of $k$, particularly $k=2$ and $k=3$, are the most suitable for detecting plagiarism generated through LLM-based code paraphrasing. 
Therefore, the answer to RQ1 is that detection performance is strongly influenced by the choice of $k$, with 2-gram birthmarks providing the best overall results among the evaluated configurations.

\subsection{RQ2: Impact of Similarity Functions}
\label{sec:rq2}

To answer RQ2, we compare the effectiveness of the evaluated similarity functions when detecting LLM-generated code clones.
Because the 1-gram birthmark demonstrated substantially lower performance in the previous analysis, it is excluded from this comparison in order to focus on practical k-gram configurations.

\begin{figure*}
    \centering
    \includegraphics[width=0.8\linewidth]{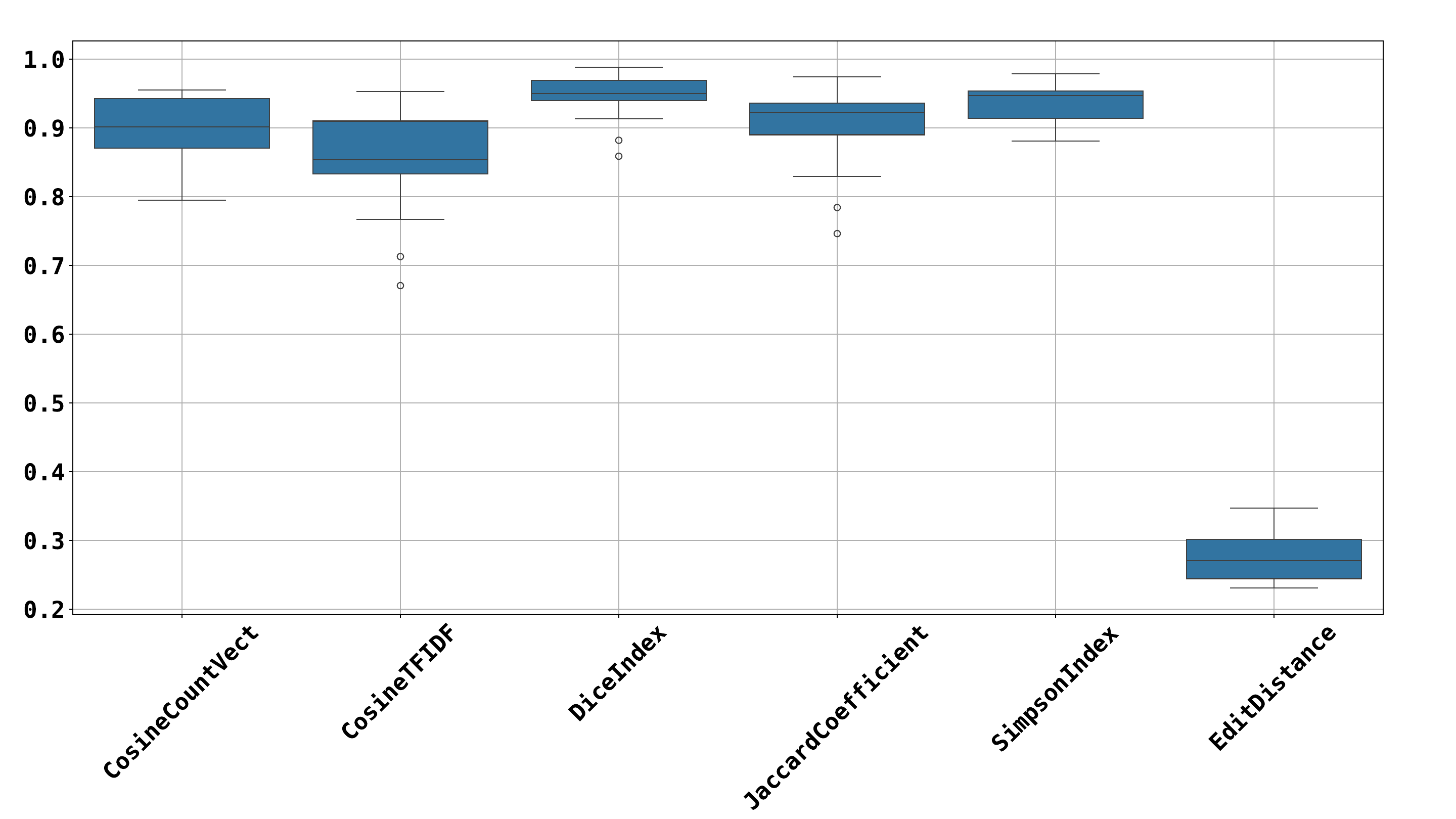}
    \caption{Comparison of similarity functions for plagiarism detection in terms of Hmean (excluding 1-gram birthmarks).}
    \label{fig:sim-func-comp}
\end{figure*}

Figure~\ref{fig:sim-func-comp} presents the detection performance of each similarity function across the evaluated birthmarks and LLM models. 
Most similarity functions achieved relatively high Hmean values, indicating that k-gram birthmarks can be effectively compared using a variety of similarity measures. 
However, edit-distance-based similarity consistently produced substantially lower performance than all other approaches.
Among the evaluated methods, the Dice index achieved the highest average performance (0.946), closely followed by the Simpson index (0.9351). 
Jaccard similarity and cosine similarity using count vectorization also produced strong results, whereas cosine similarity using TF-IDF performed slightly worse. 
In contrast, edit-distance-based similarity achieved an average Hmean of only 0.276, making it unsuitable for the considered plagiarism detection task.

One possible explanation for the poor performance of edit distance is that code paraphrasing frequently alters the ordering and arrangement of instructions. 
Although the resulting program remains functionally equivalent, modifications to control structures, statement organization, and implementation details can significantly change the sequence of extracted k-grams. 
Consequently, sequence-based comparison becomes highly sensitive to transformations introduced by the LLM.

In contrast, set-based similarity measures such as the Dice index, Jaccard coefficient, and Simpson index focus primarily on the presence of shared k-grams rather than their ordering. 
This property makes them more robust against paraphrasing-induced modifications while still preserving the ability to distinguish unrelated software. 
The superior performance of the Dice and Simpson indices suggests that emphasizing overlap between birthmark elements is particularly effective for detecting LLM-generated code clones.

Therefore, the answer to RQ2 is that the choice of similarity function has a substantial impact on detection performance. Among the evaluated methods, the Dice index provided the best overall results, while edit-distance-based similarity proved ineffective for detecting plagiarism generated through code paraphrasing.

\subsection{RQ3: Impact of LLM Selection}
\label{sec:rq3}

To answer RQ3, we investigate whether the choice of LLM affects the detectability of paraphrased code. 
For this analysis, we exclude the 1-gram birthmark and edit-distance similarity because the previous experiments demonstrated that these configurations consistently produced poor performance and would therefore dominate the comparison.

\begin{figure*}[tb]
    \centering
    \includegraphics[width=0.8\linewidth]{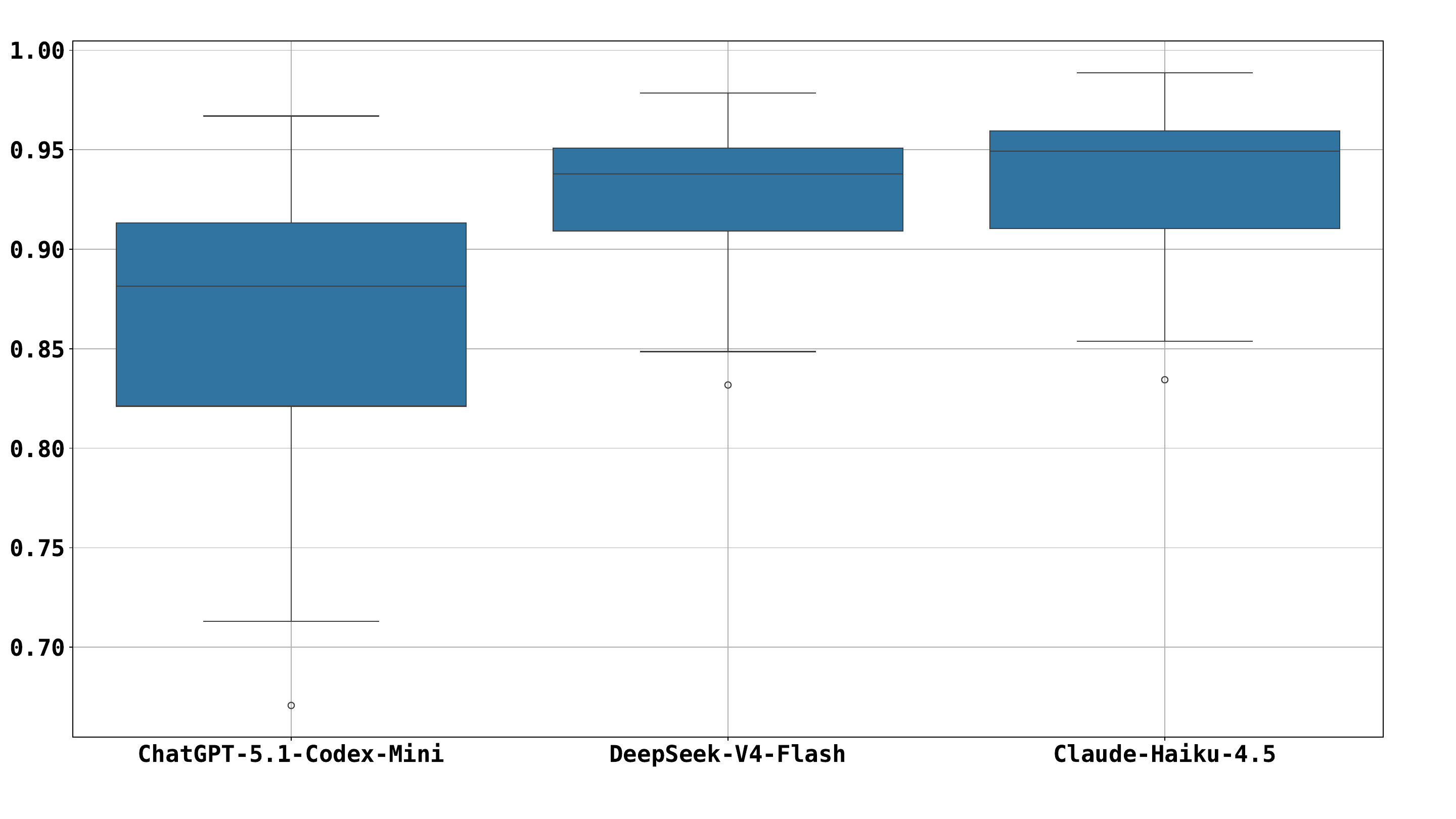}
    \caption{Impact of LLM selection on plagiarism detection performance measured using Hmean (excluding 1-gram birthmarks and edit-distance similarity).}
    \label{fig:model-comp}
\end{figure*}

Figure~\ref{fig:model-comp} presents the detection performance achieved for each evaluated model. 
Although all three models produced code clones that remained detectable by k-gram birthmarks, noticeable differences can be observed between the evaluated LLMs.

To provide a more concise comparison, Table~\ref{tab:models-avg-hmean} summarizes the average Hmean obtained for each model. 
Among the evaluated models, ChatGPT-5.1-Codex-Mini achieved the lowest average Hmean (0.8603), indicating that the paraphrased code generated by this model was generally more difficult to detect. 
In contrast, Claude-Haiku-4.5 and DeepSeek-V4-Flash achieved substantially higher Hmean values of 0.9322 and 0.9258, respectively.

\begin{table}
    \centering
    \caption{Average Hmean for different LLMs, showing the impact of model selection on plagiarism detectability (excluding 1-gram birthmarks and edit-distance similarity).}
    \label{tab:models-avg-hmean}
    \begin{tabular}{l|c}
        \toprule
        Model & Avg. Hmean \\
        \midrule
        Claude-Haiku-4.5 & 0.9322 \\
        DeepSeek-V4-Flash & 0.9258  \\
        ChatGPT-5.1-Codex-Mini & 0.8603  \\
        \bottomrule
    \end{tabular}
\end{table}

One possible explanation is that ChatGPT-5.1-Codex-Mini performs more extensive code transformations while preserving functionality, resulting in paraphrased code that differs more substantially from the original implementation. 
However, confirming this hypothesis would require a dedicated analysis of the generated code transformations and is therefore left for future work.

Interestingly, the results are consistent with the compilation statistics presented in Section~\ref{sec:compile-errors}. 
ChatGPT-5.1-Codex-Mini produced the smallest number of compilation failures while simultaneously generating the most difficult-to-detect code clones. 
This observation suggests that the model was able to apply more effective source-code transformations without sacrificing compilability.

Therefore, the answer to RQ3 is that the choice of LLM has a measurable impact on plagiarism detectability. 
Among the evaluated models, ChatGPT-5.1-Codex-Mini produced the most challenging plagiarism cases for k-gram-birthmark-based detection.

\section{Conclusion and Future Work}
\label{sec:conclusion}

In this paper, we investigated the applicability of k-gram software birthmarks to the detection of LLM-assisted code plagiarism generated through code paraphrasing. 
The experimental results provide positive evidence regarding the effectiveness of k-gram birthmarks for detecting LLM-generated paraphrased code. 
Among the evaluated birthmarks, 2-gram achieved the highest average detection performance, while 1-gram suffered from insufficient discriminative power and larger values of $k$ exhibited a gradual decline in effectiveness. 
These results suggest that short opcode sequences provide an effective balance between resilience and credibility when facing LLM-generated code transformations.

Regarding similarity functions, set-based approaches consistently produced the strongest results. 
In particular, the Dice index achieved the highest average Hmean, closely followed by the Simpson index. 
In contrast, edit-distance-based similarity performed poorly, indicating that sequence information is heavily affected by code paraphrasing and therefore unsuitable for this detection scenario.

The choice of LLM also influenced plagiarism detectability. 
Among the evaluated models, ChatGPT-5.1-Codex-Mini generated the most challenging plagiarism detection, whereas DeepSeek-V4-Flash and Claude-Haiku-4.5 produced paraphrased code that were comparatively easier to identify. 
This finding suggests that future plagiarism detection techniques should account for differences in code transformation capabilities across LLMs.

Overall, the results indicate that k-gram software birthmarks remain effective against modern LLM-based code paraphrasing, particularly when combined with small values of $k$ and set-based similarity measures. 
Given the simplicity and scalability of the approach, the findings provide a strong foundation for future research on executable-based plagiarism detection in the era of LLM-assisted software development.

As future work, we plan to extend the evaluation to additional software birthmarks and more advanced plagiarism strategies.
Furthermore, because the current dataset is limited to projects released under the BSD-2-Clause license, we intend to investigate whether the observed findings generalize to software distributed under other open-source licenses.

\medskip
\printbibliography

\medskip
\appendix 
\section{Used Prompt}
\label{sec:used-prompts}

The prompt used for generating paraphrased code. 

\begin{figure*}[h]
    \centering
    \includegraphics[width=0.8\linewidth]{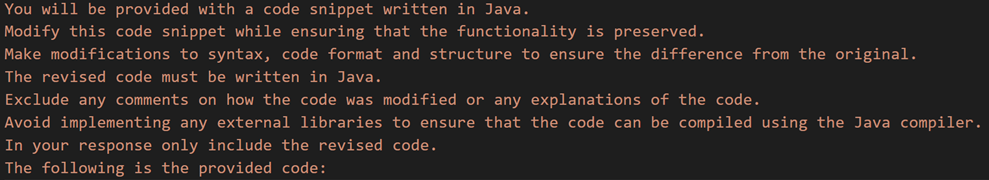}
    \label{fig:prompt-paraphrase}
\end{figure*}

\end{document}